\documentclass{PoS}
\usepackage{amsfonts,amsbsy,latexsym,amssymb,amscd,amstext}
\newcommand{\be}{\begin{equation}}
\newcommand{\ee}{\end{equation}}
\newcommand{\atan}{\mathrm{atan}}
\newcommand{\LAMMS}{\Lambda_{\overline{\rm MS}}}
\title{The string tension for Large N gauge theory from smeared Wilson loops}

\ShortTitle{The string tension for Large N gauge theory from smeared Wilson  loops}

\author{\speaker{Antonio Gonz\'alez-Arroyo}$^{ab}$ and Masanori Okawa$^c$\\
\llap{$^a$}Instituto de F\'{\i}sica Te\'orica UAM/CSIC\\
\llap{$^b$}Departamento de F\'{\i}sica Te\'orica, C-15 \\
       Universidad Aut\'onoma de Madrid, E-28049--Madrid, Spain \\
\llap{$^c$}Graduate School of Science, Hiroshima University\\
Higashi-Hiroshima, Hiroshima 739-8526, Japan\\
E-mail: \email{antonio.gonzalez-arroyo@uam.es}, \email{okawa@sci.hiroshima-u.ac.jp}}

\abstract{Using smeared Creutz ratios we extract the string tension
for SU(N) pure gauge theory and $N$=3,4,5,6,8. We employ  these results to
extrapolate to large N. The same methodology is applied to the
single-site Twisted  Eguchi Kawai model. The corresponding string
tension matches perfectly within errors with the extrapolated one,
providing strong evidence in favour of the twisted reduction
framework. Interesting results are also obtained on the behaviour of 
Creutz ratios for large sizes.}

\FullConference{The 30th International Symposium on Lattice Field Theory\\
		 June 24 - 29,  2012\\
		 Cairns, Australia}

\begin{document}

\section{Introduction}

\vspace{- 0.2 cm}

Large N gauge theories are very interesting theoretical models. They
have simpler properties than their finite N counterparts. At the
perturbative level only planar diagrams contribute. At the
non-perturbative level one has factorization, validity of the quenched
approximation, stable resonances, OZI rule, etc. Furthermore, the
connection with string theory is also simpler in the large N limit. 
On the other hand, these theories are  expected to be confining and 
have a rich meson spectrum. Thus, it is very atractive to investigate
them using lattice gauge theory methods (for a recent review see
Ref.~\cite{lucini}). The problem is that in this 
context their  simple character seems to be lost. Indeed, the standard 
pathway to obtain predictions for these theories is by extrapolating 
the results obtained for finite N. One possible simplification was 
found by Eguchi and Kawai~\cite{EK}. By examining the Migdal Makeenko
equations for Wilson loops, they concluded that, if the  $Z_N^4$
symmetry of the theory remains unbroken,  the large N pure gauge
theory on the lattice is volume independent. Bringing this idea to the
extreme they proposed a {\em reduced} single-site formulation called the
Eguchi-Kawai model.  Unfortunately, it was soon realized that in the
EK model the symmetry is broken at weak coupling~\cite{QEK}.

Since then, the question has been whether it is possible to make the 
reduction idea  survive in the continuum limit. Very soon simple 
modifications were proposed~\cite{QEK}-\cite{TEK} designed to restore
the $Z^4_N$ symmetry of the EK model.  Recently new proposals have been 
added to the list~\cite{KUY}-\cite{Shifman}.  Alternatively, Narajanan and
Neuberger~\cite{NaNeu}
proposed a mixed procedure called {\em partial reduction}, and
suggested that presumably reduction only operates beyond the
deconfinement scale. 

Our goal is to test whether the twisted reduction proposal, introduced
by the present authors~\cite{TEK}, is indeed capable of realizing the  
reduction  idea in the continuum limit. This is based on the Twisted
Eguchi Kawai model (TEK), a single site model whose action is 
given by
\be
\label{actionTEK}
S= bN \sum_{\mu\ne \nu} \mathrm{Tr}\left( z_{\mu \nu} U_\mu U_\nu U_\mu^\dagger
U_\nu^\dagger\right)
\ee
where $U_\mu$ are SU(N) matrices,  $b$ is the inverse `t Hooft coupling and $z_{\mu \nu}$ are
elements of the center $Z_N$. The choice of $z_{\mu \nu}$ is crucial
for determining the behaviour of the system in the weak coupling limit
(large $b$). Choosing $z=1$ gives the original EK model, which has an
infinite degeneracy at the classical level~\cite{groundstate}.
Although the action is invariant under $Z_N^4$ symmetry ($U_\mu
\longrightarrow z_\mu U_\mu$), the quantum corrections favour
symmetry-breaking minima. 

Other choices of $z_{\mu \nu}$ lift the degeneracy of the ground state
and the classical vacuum becomes invariant under a subgroup of $Z_N^4$. 
A particularly elegant choice is termed {\em symmetric twist} and
given by ($\nu > \mu$)
\be
z_{\mu \nu}=z^*_{\nu \mu}=\exp\{ 2 \pi i \frac{k}{L}\}
\ee
where $N=L^2$ and $k$ is an integer defined modulo $L$. With this
choice, the classical vacuum is invariant under a $Z_L^4$ subgroup of the 
$Z_N^4$. In the large L limit (large N) this is enough to secure the 
reduction idea at weak coupling. Perturbative expansion around this
vacuum gives rise to modified Feynman rules. The propagators become
just those of a lattice field theory at finite volume $L^4$. This is an 
important information showing how the $N^2$ degrees of freedom of the
group map fully onto effective spatial degrees of freedom. It also
suggests what are the dominant finite $N$ corrections to reduction. 
The remaining Feynman rules introduce momentum dependence at the vertices,
providing a  discretized version of {\em non-commutative field theory}.

Most of the early studies of the TEK model were done for $k=1$ and
showed that the $Z_L^4$ symmetry remained unbroken at intermediate
couplings. A few years ago it was found~\cite{IO}-\cite{TV}-\cite{AHHI} 
that at intermediate couplings
the symmetry breaks down for $N$>$N_c$=100. The critical value of $N$ depends
on $k$  roughly as  $N_c(k) \sim 90 k^2$. Thus, in Ref.~\cite{agamo} we
proposed that the correct large $N$ limit has to be taken keeping $k/L$
approximately fixed. Our goal is to test this idea and use the reduced model to
obtain physical information from the large N continuum gauge theory at infinite
volume. We are currently involved in this task and the present talk
reports part of these results. 

It must be said that we have been able to simulate the model up to
$N$=1369 without finding any indication of symmetry breaking. Notice
that this amounts to effective lattice sizes of $37^4$ which are
normally considered large enough in ordinary lattice gauge theory
simulations. To try to understand the interplay between finite $N$ and
finite volume in a more detailed form we have analysed in certain
detail the 3 dimensional version of the theory. Some results from this
work,
done in collaboration with Margarita Garc\'{\i}a P\'erez, has been
reported by her at this conference~\cite{marga}. The results support
our claims that, if the flux $k$ is chosen judiciously, the physical results
essentially depend on $Nl$, the  product of torus linear size $l$ and 
group rank. This is indeed consistent with the volume independence
achieved at infinite $N$. 

The present work reports another form of evidence in favour of the 
twisted reduction idea. We set ourselves the goal of computing the 
continuum string tension at large N using the reduced model. To make
the comparison free of systematics associated with the methodology, 
we decided to use exactly the same procedure to extract the string
tension for ordinary SU(N) lattice gauge theory, and then  extrapolate
these results  to infinite $N$. The main 
results have appeared~\cite{agamotwo} recently. Here we will 
show  our updated  results which include those for SU(4).

\vspace{- 0.3 cm}

\section{Numerical results on the string tension}

\vspace{- 0.2 cm}

The basic data is made up several independent lattice gauge theory
simulations for different values of $N$ and lattice gauge coupling $b$
and a spatial volume of $32^4$. Each simulation contains 260
configurations separated by 100 sweeps formed by 1 heat-bath and 5
overrelaxation updates, after discarding 4000 sweeps to reach
thermalization.  We simulated the SU(N) gauge groups for $N$=3,4, 5, 6
and 8. For each group we made independent simulations at different
values of $b$ (7 for $N$=3 and 4, and 5 for $N$=5,6,8). This huge
information can be used to determine the string tension and to
extrapolate to the continuum limit. 

In addition, we  made simulations of the TEK model  for $N=29^2=841$,
$k=9$
and 7 values of $b$. The number of configurations in that case was
around 6000 with similar updating procedure and separation of sweeps. 

In all cases we computed $T\times R$ rectangular Wilson loops $W(T,R)$ computed
from smeared links, and from them we extracted the Creutz ratios
defined as 
\begin{equation}
\chi(T,R)= -\log \frac{W(T+0.5,R+0.5)W(T-0.5,R-0.5)}{W(T+0.5,R-0.5)W(T-0.5,R+0.5)}
\end{equation}
Errors were computed by jack-knife.  Details of the procedure will be
given elsewhere~\cite{agamotwo}-\cite{agamothree}. Here we will focus upon the results.

Our goal is to obtain the continuum string tension for all gauge
groups. This is a two step procedure in which one first obtains 
the lattice string tension as 
$\kappa=\lim_{R,T\longrightarrow \infty}\chi(T,R)$, 
and then one makes a scaling analysis of the
results to take the continuum limit. The first part can be done by 
fitting the square Creutz ratio data (R=T) to the formula
\begin{equation}
\chi(R,R)= \kappa+ \frac{2 \gamma}{R^2}+\frac{\eta}{R^4} 
\label{firstfit}
\end{equation}
Different ranges have been examined, and order 1  chi squares per degree
of freedom are achieved using the data $R=T\in [3.5,8.5]$. As an example, 
we show in Fig.~1 one such fit. 

\begin{figure}
\label{figA}
\centering
\includegraphics[width=.6\textwidth]{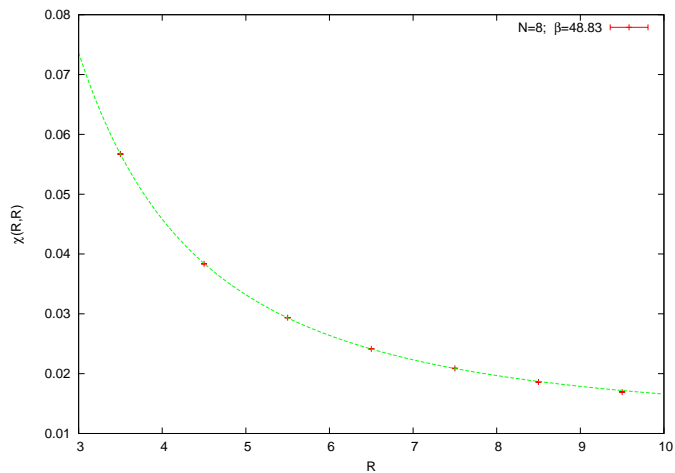}
\caption{$\chi (R,R)$ compared to the best fit
Eq.~(2.2)
for $N$=8 and b=0.3815}
\end{figure}

Once the  different values of $\kappa(N,b)$ are extracted from the
fits, one must perform a scaling analysis. This is done by taking the
limit 
\begin{equation}
\sigma(N)= \lim_{a(N,b) \longrightarrow 0} \, \frac{\kappa(N,b)}{a^2(N,b)} 
\end{equation}
where $a(N,b)$ gives the lattice spacing (in certain units) as a
function of the coupling $b$. Different choices of $a(N,b)$ correspond
to different renormalization schemes. A good choice makes the approach
to the $a=0$ limit  smoother. We have tried several expressions found in the
literature for $a(N,b)$. In all cases, the approach seems to be linear 
in $a^2$ with slopes depending on the scheme and  on the rank of the
group $N$. A class of schemes uses perturbation theory and has the
advantage of expressing the lattice spacing in units of $\LAMMS$. We
choose here the choice taken in Ref.~\cite{allton} which has been
tested and used for similar purposes as ours. 

Finally, the resulting string tensions $\sigma(N)$ are studied  as a
function of  $N$ and extrapolated to large $N$. In Fig.~2 
we display the ratio $\LAMMS/\sqrt{\sigma(N)}$ as a
function of $1/N^2$. The data shows a nice linear
pattern that allows an extrapolation to infinite $N$. The best fit
value gives $\LAMMS/\sqrt{\sigma(\infty)}=0.525(2)$. The result  obtained from
the TEK model is  $\LAMMS/\sqrt{\sigma(\infty)}=0.523(5)$. The
agreement is very
good and provides a nice test that  reduction is operative in the
continuum  within the confinement regime of the theory. The $1/N^2$
dependence displayed by the data matches perfectly with the result 
obtained in Ref.~\cite{allton}. Our estimate of $\sigma(\infty)$,
however, disagrees with the one given in that reference within
statistical errors, although not if the systematic errors given in
Ref.~\cite{allton} are taken into account. A new estimate 
of the large N string tension has appeared recently~\cite{Lohmayer}. 
The methodology is also based upon smeared Wilson loops but the volumes
and values of $N$ used  sit in between our conventional and fully
reduced model. It is interesting  to point out that their result is in
agreement with ours and differs from the best estimate of
Ref.~\cite{allton}.

\begin{figure}
\label{figB}
\begingroup
  \makeatletter
  \providecommand\color[2][]{%
    \GenericError{(gnuplot) \space\space\space\@spaces}{%
      Package color not loaded in conjunction with
      terminal option `colourtext'%
    }{See the gnuplot documentation for explanation.%
    }{Either use 'blacktext' in gnuplot or load the package
      color.sty in LaTeX.}%
    \renewcommand\color[2][]{}%
  }%
  \providecommand\includegraphics[2][]{%
    \GenericError{(gnuplot) \space\space\space\@spaces}{%
      Package graphicx or graphics not loaded%
    }{See the gnuplot documentation for explanation.%
    }{The gnuplot epslatex terminal needs graphicx.sty or graphics.sty.}%
    \renewcommand\includegraphics[2][]{}%
  }%
  \providecommand\rotatebox[2]{#2}%
  \@ifundefined{ifGPcolor}{%
    \newif\ifGPcolor
    \GPcolorfalse
  }{}%
  \@ifundefined{ifGPblacktext}{%
    \newif\ifGPblacktext
    \GPblacktexttrue
  }{}%
  \let\gplgaddtomacro\g@addto@macro
  \gdef\gplbacktext{}%
  \gdef\gplfronttext{}%
  \makeatother
  \ifGPblacktext
    \def\colorrgb#1{}%
    \def\colorgray#1{}%
  \else
    \ifGPcolor
      \def\colorrgb#1{\color[rgb]{#1}}%
      \def\colorgray#1{\color[gray]{#1}}%
      \expandafter\def\csname LTw\endcsname{\color{white}}%
      \expandafter\def\csname LTb\endcsname{\color{black}}%
      \expandafter\def\csname LTa\endcsname{\color{black}}%
      \expandafter\def\csname LT0\endcsname{\color[rgb]{1,0,0}}%
      \expandafter\def\csname LT1\endcsname{\color[rgb]{0,1,0}}%
      \expandafter\def\csname LT2\endcsname{\color[rgb]{0,0,1}}%
      \expandafter\def\csname LT3\endcsname{\color[rgb]{1,0,1}}%
      \expandafter\def\csname LT4\endcsname{\color[rgb]{0,1,1}}%
      \expandafter\def\csname LT5\endcsname{\color[rgb]{1,1,0}}%
      \expandafter\def\csname LT6\endcsname{\color[rgb]{0,0,0}}%
      \expandafter\def\csname LT7\endcsname{\color[rgb]{1,0.3,0}}%
      \expandafter\def\csname LT8\endcsname{\color[rgb]{0.5,0.5,0.5}}%
    \else
      \def\colorrgb#1{\color{black}}%
      \def\colorgray#1{\color[gray]{#1}}%
      \expandafter\def\csname LTw\endcsname{\color{white}}%
      \expandafter\def\csname LTb\endcsname{\color{black}}%
      \expandafter\def\csname LTa\endcsname{\color{black}}%
      \expandafter\def\csname LT0\endcsname{\color{black}}%
      \expandafter\def\csname LT1\endcsname{\color{black}}%
      \expandafter\def\csname LT2\endcsname{\color{black}}%
      \expandafter\def\csname LT3\endcsname{\color{black}}%
      \expandafter\def\csname LT4\endcsname{\color{black}}%
      \expandafter\def\csname LT5\endcsname{\color{black}}%
      \expandafter\def\csname LT6\endcsname{\color{black}}%
      \expandafter\def\csname LT7\endcsname{\color{black}}%
      \expandafter\def\csname LT8\endcsname{\color{black}}%
    \fi
  \fi
  \setlength{\unitlength}{0.0500bp}%
  \begin{picture}(7200.00,5040.00)%
    \gplgaddtomacro\gplbacktext{%
      \csname LTb\endcsname%
      \put(1078,704){\makebox(0,0)[r]{\strut{} 0.51}}%
      \put(1078,1518){\makebox(0,0)[r]{\strut{} 0.52}}%
      \put(1078,2332){\makebox(0,0)[r]{\strut{} 0.53}}%
      \put(1078,3147){\makebox(0,0)[r]{\strut{} 0.54}}%
      \put(1078,3961){\makebox(0,0)[r]{\strut{} 0.55}}%
      \put(1078,4775){\makebox(0,0)[r]{\strut{} 0.56}}%
      \put(1640,484){\makebox(0,0){\strut{} 0}}%
      \put(2501,484){\makebox(0,0){\strut{} 0.02}}%
      \put(3361,484){\makebox(0,0){\strut{} 0.04}}%
      \put(4222,484){\makebox(0,0){\strut{} 0.06}}%
      \put(5082,484){\makebox(0,0){\strut{} 0.08}}%
      \put(5943,484){\makebox(0,0){\strut{} 0.1}}%
      \put(6803,484){\makebox(0,0){\strut{} 0.12}}%
      \put(176,2739){\rotatebox{-270}{\makebox(0,0){\strut{}$\frac{\Lambda_{\overline{MS}}}{\sqrt{\sigma}}$}}}%
      \put(4006,154){\makebox(0,0){\strut{}$1/N^{2}$}}%
    }%
    \gplgaddtomacro\gplfronttext{%
      \csname LTb\endcsname%
      \put(5816,1317){\makebox(0,0)[r]{\strut{}SU(N)}}%
      \csname LTb\endcsname%
      \put(5816,1097){\makebox(0,0)[r]{\strut{}TEK N=841}}%
      \csname LTb\endcsname%
      \put(5816,877){\makebox(0,0)[r]{\strut{}0.5256+0.255/$N^{2}$}}%
    }%
    \gplbacktext
    \put(0,0){\includegraphics{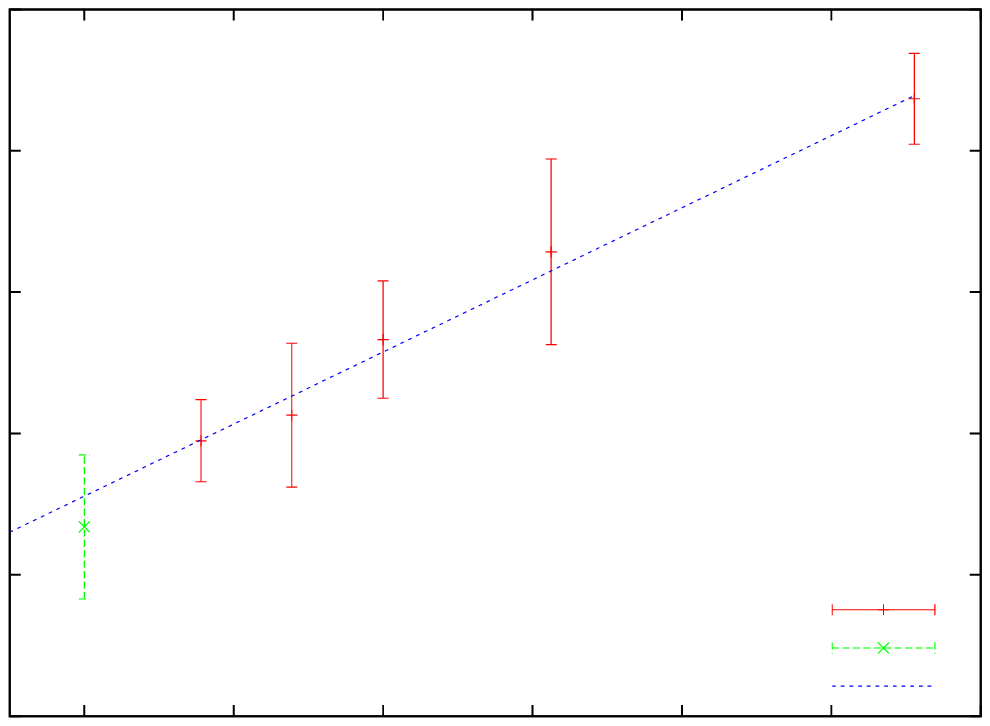}}%
    \gplfronttext
  \end{picture}%
\endgroup
\caption{$\LAMMS/\sqrt{\sigma}$ as a funtion of 1/$N^2$}
\end{figure}

We believe that the main source of systematic error in our estimate of 
$\LAMMS/\sqrt{\sigma(\infty)}$ is indeed sitting in the scale itself.
For example, using other perturbative schemes one gets differences of
the order of two percent. For that purpose it is better to use a
non-perturbative renormalization scheme based on the data itself. This
can be achieved by inverting the order of the limits $R\longrightarrow
\infty$ and $a\longrightarrow 0$. That means that we will take the
continuum limit of the Creutz ratios. This is possible since   
Creutz ratios  are free from corner and perimeter divergences. Notice,
however,  that   Creutz ratios have intrinsic scaling violations due
to its mere definition. Thus, in the continuum limit one should have 
\be
\chi(T,R)= a^2(b) \tilde{F}(t,r) + a^4(b)  \tilde{H}(t,r)+ \ldots
\ee
where $t=Ta(b)$, $r=Ra(b)$ and $\tilde{F}(t,r)$ and $\tilde{H}(t,r)$
are well-defined continuum functions.  Indeed,  $\tilde{F}(t,r)$ is
 given by 
\be
\label{ftilde}
\tilde{F}(t,r)=-\frac{\partial^2 \log({\cal W}(t,r))}{\partial r
\partial t}= \sigma + \gamma(r/t) \left(
\frac{1}{r^2}+\frac{1}{t^2}\right) +\ldots
\ee
where ${\cal W}(t,r)$ is the continuum Wilson loop (divergences drop
from the formula). The expression on the right-hand side is the
asymptotic prediction from an effective string theory description of 
the Wilson loop expectation value. Indeed, the Nambu-Goto theory
predicts 
\be
\gamma_{NG}(z)=-\frac{1}{1+1/z^2} \frac{\partial}{\partial
z}\left( z \frac{\partial \log(\eta(iz))}{\partial
 z}\right) 
\ee
where $\eta(iz)$ is the Dedeking eta function. Indeed, perturbation
theory is also compatible with the form given in Eq.~\ref{ftilde}, with  
$\sigma=0$ and
\be
\gamma_{PT}(z)=\frac{(N^2-1) g^2}{4 \pi^2 N} \frac{1+z \atan(z)+ \atan(1/z)/z}{z+1/z}
\ee

The continuum function $\tilde{F}(t,r)$ can be used to fix a physical scale 
$\bar{r}$ as follows:
\be
\bar{r}^2  \tilde{F}(\bar{r},\bar{r})=1.65
\ee
The choice of 1.65 is conventional and suggested by the analogy
between our scheme and the so-called Sommer scale. 

Once the scale is fixed, one can use the data to obtain a determination
of $\tilde{F}(r,t)$  for each $N$ using all the Creutz ratios for all  
values of $R$, $T$ and $b$. In particular, our result for  $\bar{r}^2  \tilde{F}(r,r)$
and $N$=8 is given in  Fig.~3. The data are very well fitted 
to by a second degree polynomial in $\frac{\bar{r}^2}{r^2}$ with a
fairly small quadratic term. Similar behaviour applies for other
values of $N$ and for the TEK model. From the data we estimate 
$\sigma(\infty) \bar{r}^2= 1.105(10)$ and $\gamma(1)=0.272(5)$. The
latter value differs from the prediction of Nambu-Goto theory
$\gamma_{NG}(1)=0.16{\ldots}$. Similarly, we analysed the behaviour of 
$\gamma(z)$ for $z$ close to one. The data is well described by a 
parametrization $\gamma(z)=\gamma(1)(1+ \tau \frac{(z-1)^2}{2z})$. 
with $\tau=0.31(6)$. This value is fairly close to perturbative result
0.39   obtained from $\gamma_{PT}(z)/\gamma_{PT}(1)$, and differs  significantly
 from  the value $\sim 2$ obtained from $\gamma_{NG}$. The result
 might indicate that one has to supplement the string contribution
 with one coming from one gluon exchange.

\begin{figure}
\label{figC}
\begingroup
  \makeatletter
  \providecommand\color[2][]{%
    \GenericError{(gnuplot) \space\space\space\@spaces}{%
      Package color not loaded in conjunction with
      terminal option `colourtext'%
    }{See the gnuplot documentation for explanation.%
    }{Either use 'blacktext' in gnuplot or load the package
      color.sty in LaTeX.}%
    \renewcommand\color[2][]{}%
  }%
  \providecommand\includegraphics[2][]{%
    \GenericError{(gnuplot) \space\space\space\@spaces}{%
      Package graphicx or graphics not loaded%
    }{See the gnuplot documentation for explanation.%
    }{The gnuplot epslatex terminal needs graphicx.sty or graphics.sty.}%
    \renewcommand\includegraphics[2][]{}%
  }%
  \providecommand\rotatebox[2]{#2}%
  \@ifundefined{ifGPcolor}{%
    \newif\ifGPcolor
    \GPcolortrue
  }{}%
  \@ifundefined{ifGPblacktext}{%
    \newif\ifGPblacktext
    \GPblacktexttrue
  }{}%
  \let\gplgaddtomacro\g@addto@macro
  \gdef\gplbacktext{}%
  \gdef\gplfronttext{}%
  \makeatother
  \ifGPblacktext
    \def\colorrgb#1{}%
    \def\colorgray#1{}%
  \else
    \ifGPcolor
      \def\colorrgb#1{\color[rgb]{#1}}%
      \def\colorgray#1{\color[gray]{#1}}%
      \expandafter\def\csname LTw\endcsname{\color{white}}%
      \expandafter\def\csname LTb\endcsname{\color{black}}%
      \expandafter\def\csname LTa\endcsname{\color{black}}%
      \expandafter\def\csname LT0\endcsname{\color[rgb]{1,0,0}}%
      \expandafter\def\csname LT1\endcsname{\color[rgb]{0,1,0}}%
      \expandafter\def\csname LT2\endcsname{\color[rgb]{0,0,1}}%
      \expandafter\def\csname LT3\endcsname{\color[rgb]{1,0,1}}%
      \expandafter\def\csname LT4\endcsname{\color[rgb]{0,1,1}}%
      \expandafter\def\csname LT5\endcsname{\color[rgb]{1,1,0}}%
      \expandafter\def\csname LT6\endcsname{\color[rgb]{0,0,0}}%
      \expandafter\def\csname LT7\endcsname{\color[rgb]{1,0.3,0}}%
      \expandafter\def\csname LT8\endcsname{\color[rgb]{0.5,0.5,0.5}}%
    \else
      \def\colorrgb#1{\color{black}}%
      \def\colorgray#1{\color[gray]{#1}}%
      \expandafter\def\csname LTw\endcsname{\color{white}}%
      \expandafter\def\csname LTb\endcsname{\color{black}}%
      \expandafter\def\csname LTa\endcsname{\color{black}}%
      \expandafter\def\csname LT0\endcsname{\color{black}}%
      \expandafter\def\csname LT1\endcsname{\color{black}}%
      \expandafter\def\csname LT2\endcsname{\color{black}}%
      \expandafter\def\csname LT3\endcsname{\color{black}}%
      \expandafter\def\csname LT4\endcsname{\color{black}}%
      \expandafter\def\csname LT5\endcsname{\color{black}}%
      \expandafter\def\csname LT6\endcsname{\color{black}}%
      \expandafter\def\csname LT7\endcsname{\color{black}}%
      \expandafter\def\csname LT8\endcsname{\color{black}}%
    \fi
  \fi
  \setlength{\unitlength}{0.0500bp}%
  \begin{picture}(8220.00,4818.00)%
    \gplgaddtomacro\gplbacktext{%
      \csname LTb\endcsname%
      \put(726,704){\makebox(0,0)[r]{\strut{} 1}}%
      \put(726,1474){\makebox(0,0)[r]{\strut{} 1.5}}%
      \put(726,2244){\makebox(0,0)[r]{\strut{} 2}}%
      \put(726,3013){\makebox(0,0)[r]{\strut{} 2.5}}%
      \put(726,3783){\makebox(0,0)[r]{\strut{} 3}}%
      \put(726,4553){\makebox(0,0)[r]{\strut{} 3.5}}%
      \put(858,484){\makebox(0,0){\strut{} 0}}%
      \put(1729,484){\makebox(0,0){\strut{} 1}}%
      \put(2599,484){\makebox(0,0){\strut{} 2}}%
      \put(3470,484){\makebox(0,0){\strut{} 3}}%
      \put(4341,484){\makebox(0,0){\strut{} 4}}%
      \put(5211,484){\makebox(0,0){\strut{} 5}}%
      \put(6082,484){\makebox(0,0){\strut{} 6}}%
      \put(6952,484){\makebox(0,0){\strut{} 7}}%
      \put(7823,484){\makebox(0,0){\strut{} 8}}%
      \put(4340,154){\makebox(0,0){\strut{}$2 \frac{\bar{r}^2}{r^2}$}}%
    }%
    \gplgaddtomacro\gplfronttext{%
      \csname LTb\endcsname%
      \put(2706,4380){\makebox(0,0)[r]{\strut{}$\bar{r}^2 \tilde{F}(r,r)$  }}%
      \csname LTb\endcsname%
      \put(2706,4160){\makebox(0,0)[r]{\strut{}Linear part}}%
      \csname LTb\endcsname%
      \put(2706,3940){\makebox(0,0)[r]{\strut{}Quadratic fit}}%
    }%
    \gplbacktext
    \put(0,0){\includegraphics{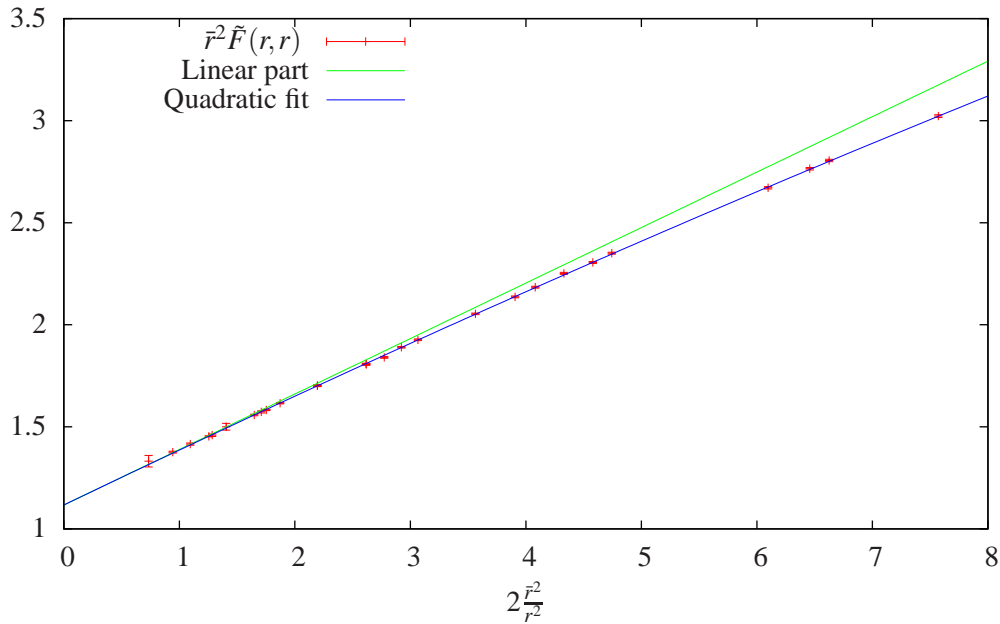}}%
    \gplfronttext
  \end{picture}%
\endgroup
\caption{ The continuum function  $\bar{r}^2  \tilde{F}(r,r)$ for $N$=8}
\end{figure}

\vspace{- 0.1 cm}

\section{Conclusions}

\vspace{- 0.1 cm}

The main conclusions of our work are:

$\bullet$ We have presented a new method to determine the string tension
from smeared Creutz ratios.

$\bullet$ The results scale smoothly to the continuum limit.

$\bullet$  The continuum string tension  extrapolates linearly in
$1/N^2$ towards the large N limit. The result matches with that
obtained from the TEK reduced model. This gives a strong support for
the validity of continuum reduction.

$\bullet$  The data satisfy nice scaling properties, that allow  the
reconstruction of a finite continuum function of Wilson loops, called
$\tilde{F}(r,t)$.

$\bullet$ The function gives rise to the definition of a new non-perturbative
renormalization scheme, and accompanying scale ${\bar r}$.

$\bullet$  The subleading behaviour of  $\tilde{F}(r,t)$  differs
from the prediction of Nambu-Goto theory.

\vspace{  0.2 cm}

A.G-A is supported from Spanish grants FPA2009-08785, FPA2009-09017, CSD2007-00042, HEPHACOS S2009/ESP-1473,
PITN-GA-2009-238353 (ITN STRONGnet) and CPAN CSD2007-00042. M.O is supported in part by Grants-in-Aid for Scientific Research from the Ministry of Education, Culture, Sports, Science and Technology (No 23540310).

The calculation has been done on Hitachi SR16000-M1 computer at High Energy Accelerator
Research Organization (KEK) supported by the Large Scale Simulation Program No.12-01
(FY2011-12).

\end{document}